\documentclass{article}

\usepackage{arxiv}

\usepackage[utf8]{inputenc} 
\usepackage[T1]{fontenc}    
\usepackage{hyperref}       
\usepackage{url}            
\usepackage{booktabs}       
\usepackage{amsmath,amsfonts,amssymb}       
\usepackage{nicefrac}       
\usepackage{microtype}      
\usepackage{graphicx}
\usepackage[backend=bibtex, style=numeric, sorting=none, citestyle=numeric]{biblatex}
\usepackage{doi}

\title{Meta-Optics Triplet for Zoom Imaging at Mid-Wave Infrared}

\author{Anna Wirth-Singh$^{1,*}$\\
\And
Arturo Martin Jimenez$^{2}$
\And
Minho Choi$^{3}$
\And
Johannes E. Fr\"{o}ch$^{1,3}$\\
\And 
Rose Johnson$^{3}$
\And
Tina Le Teichmann$^{1}$
\And
Zachary Coppens$^{2}$
\And
Arka Majumdar$^{1,3,**}$\\
\And
\\
$^{1}${Department of Physics, University of Washington, Seattle, WA, USA}\\
$^{2}${CFD Research Corporation, Huntsville, AL, USA}\\
$^{3}${Department of Electrical and Computer Engineering, University of Washington, Seattle, WA, USA}\\
$^{*}$\textit{annaw77@uw.edu}\\
$^{**}$\textit{arka@uw.edu}
}

\begin{document}
\maketitle

\begin{abstract}
Lenses with dynamic focal length, also called zoom functionality, enable a variety of applications related to imaging and sensing. The traditional approach of stacking refractive lenses to achieve this functionality results in an expensive, heavy optical system. Especially for applications in the mid-infrared, light weight and compact form factor are required. In this work, we use a meta-optic triplet to demonstrate zoom imaging at mid-wave infrared wavelengths. By varying the axial distances between the optics, the meta-optic triplet achieves high quality imaging over a zoom range of 5x, with 50$^\circ$ full field of view in the widest configuration and an aperture of 8 mm. This triplet system demonstrates the potential for meta-optics to reduce conventional components in complex and multi-functional imaging systems to dramatically thinner and lighter components. 
\end{abstract}

\section{\label{sec:Introduction}Introduction}

Thermal imaging has extensive applications in defense, night vision, and meteorology \cite{Wei23,Sanson10,Wilson2023,Estrera03}. In particular, the mid-wave infrared (MWIR, 3-5 $\mu$m) is especially attractive for long-range imaging and sensing due to its superior penetration in high-humidity atmospheric conditions \cite{Dhar08}. Wide field of view and high resolution are often desirable for these imaging applications. For imaging dynamic targets, it is particularly helpful to have both wide field of view images for context and narrow field of view for zoomed-in detail, with rapid switching mechanism between the views. However, achieving both qualities simultaneously requires a large, high-resolution camera sensor, and such sensors are particularly expensive at MWIR wavelengths. Alternatively, these applications can benefit from zoom functionality, wherein the image magnification is changed by repositioning optics within the lens system so that a range of image magnifications can be captured using a single aperture and sensor. 

Most conventional zoom systems are composed of a series of refractive lenses, some of which are axially translated to change the zoom state. While effective, stacked refractive lenses quickly become heavy, which is undesirable for many applications where zoom functionality is required, such as airborne scopes and cameras. Due to the long total track length of zoom imaging systems, excess weight can cause significant torque on the mechanical support system. In addition, to achieve high quality imaging with zoom constraints, aspheric lenses are often used \cite{Zhang24,Wei23} that are difficult to manufacture due to non-constant curvature across the surface. Combined with the limited availability of suitable materials for MWIR refractive lenses, such as germanium and calcium fluoride, this contributes to the increased cost of thermal zoom imaging systems.

Ultra-thin meta-optics have demonstrated great potential to miniaturize imaging systems, including those at MWIR wavelengths \cite{Shalaginov20,Yue23,Shih22}. However, introducing zoom functionality into meta-optics, and other forms of metasurfaces, is challenging. Various techniques have been employed, including mechanical stretching \cite{Ee16,Kamali16,She18}, MEMS \cite{Arbabi18}, and Alvarez \cite{Zhan17,Colburn18} lenses, but these techniques are limited in zoom range and aperture. Similarly to the conventional approach of stacking a series of refractive lenses, meta-optics can also be stacked to improve imaging performance. Meta-optic doublets for applications including aberration correction \cite{Arba16,Groever17} and large field of view \cite{Park23} have been demonstrated, but without a tuning mechanism. Recently, researchers have demonstrated zoom imaging by axial translation in a doublet lens system at 940 nm using two meta-optics with extended depth of focus for extended zoom range, but for a relatively small aperture of 1 mm \cite{Zhang24}. Further extension into triplet meta-optic systems is a relatively unexplored area, with existing demonstrations focused mainly on chromatic aberration correction  \cite{Shrestha23,Pan23}.

\begin{figure}[h!]
\centering\includegraphics[width=14cm]{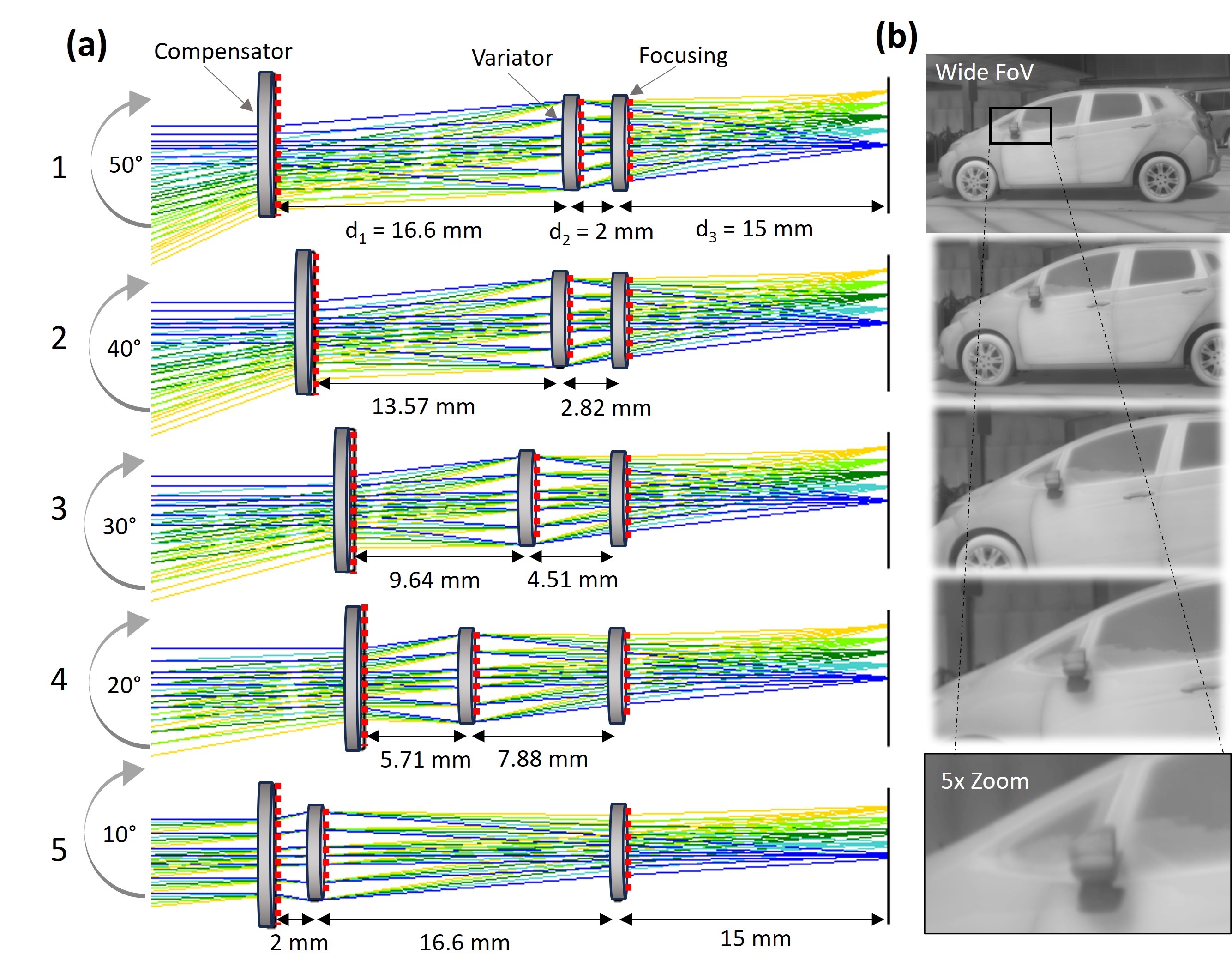}
\caption{\label{fig:Schematic}The zoom lens concept. (a) Ray tracing diagrams for all five configurations. By translating the compensator and variator optics, the effective focal length of the system is changed from wide field of view (top) to narrow field of view (bottom). The blue lines represent normally incident rays, while the yellow rays represent $25^\circ$ and $5^\circ$ angle of incidence for the wide and narrow configurations, respectively. Intermediate angles are shown in shades of green. (b) Illustration of zoom imaging on a car. Under 5x zoom, the details of the side mirror and windows become clear. }
\end{figure}

In this work, we demonstrate an 8 mm aperture, 50$^\circ$ full field of view (FoV) zoom imaging meta-optic triplet at 3.4 $\mu$m wavelength. We follow the traditional approach by varying the position of two of the three meta-optic elements to increase the image magnification from 1x to 5x. Additionally, our zoom imaging system is parfocal, meaning that the focal plane remains fixed while achieving the intermediate zoom states. For optimal transmission through the triplet lens system, we realize the meta-optics in a silicon-on-sapphire material platform. We demonstrate high-quality imaging performance up to 50$^\circ$ FoV in the wide field of view configuration.  This development enables MWIR applications with zoom functionality and lightweight multi-element imaging systems.

\section{Zoom Lens Design}

At minimum, most optical zoom systems consist of a mechanically adjustable afocal frontend and imaging backend. In order to keep a fixed back focal length, the front part of the system must be composed of two moving groups \cite{Sanson10}, resulting in a triplet lens system. The design for the meta-optic zoom lens follows this standard three-group zoom structure (consisting of a compensator, a variator, and a focusing optics) and the zoom behavior is schematically depicted in Figure 1. The system is optimized for five discrete zoom states, denoted configurations 1 through 5, with this change accomplished by varying the axial distances between the metasurfaces. Specifically, the meta-optic nearest the sensor, termed the focusing optic, remains a fixed 15 mm distance from the sensor while the positions of the other two optics (denoted as the compensator optic nearest the object and the variator optic in the middle) are adjusted to change the zoom state. This change is primarily accomplished by axially translating the variator optic, which is adjusted within a range of 14.6 mm. The compensator optic has negative power to defocus the light and its axial position is adjusted within a smaller range of 5.21 mm to correct for aberrations. Ray tracing diagrams of all configurations are shown in Figure \ref{fig:Schematic}a. 

 The designed F-number, effective focal length (EFL), and full field of view of each configuration are summarized in Table \ref{Tab:Configurations}. Configuration 1, also called the "wide" configuration, provides the widest field of view at $50^\circ$ full FoV and F-number of 3.7. Configuration 5, also called the "tele" configuration, provides the most zoomed-in image with $10^\circ$ full FoV at an F-number of 8.0. The intermediate Configurations 2 through 4 provide equally spaced intermediate zoom states. 
 
\begin{table}[h]
\caption{System Design Specifications}\label{Tab:Configurations}
\begin{tabular*}{\textwidth}{@{\extracolsep\fill}cccccc}
\toprule
 Configuration & 1 (Wide)  & 2 &  3 (Mid)  &  4 &  5 (Tele) \\
\midrule
F Number & 3.71 & 3.94 & 4.40 & 5.33 & 7.98 \\
EFL (mm) & 8.78 & 10.53 & 14.19 & 21.74 & 43.89 \\
Full FoV & 50$^\circ$ & 40$^\circ$ & 30$^\circ$ & 20$^\circ$ & 10$^\circ$ \\
\midrule
$d_1$ (mm) & 16.60 & 13.57 & 9.64 & 5.71 & 2.00  \\
$d_2$ (mm) & 2.00 & 2.82 & 4.51 & 7.88  & 16.60 \\
\bottomrule
\end{tabular*}
\end{table}

The phase profiles of the meta-optics were optimized using commercial ray tracing software (Zemax OpticStudio). Each meta-optic was modeled as a Binary-2 phase profile according to the equation
\begin{equation}
\Phi(\rho) = \sum_{i=1}^{6} A_{i} \left( \frac{\rho}{M} \right) ^{2i}    
\end{equation}\label{eqn:Binary2}
where $M$ is a normalization constant, $\rho$ is the radial coordinate, and $A_{i}$ are polynomial coefficients. To optimize the zoom system, we initially varied the phase profile coefficients $A_{i}$ to optimize performance at both the tele and wide configurations at predetermined positions. Specifically, wavefront error (with respect to an ideal spherical wavefront) was defined as the merit function during optimization. The optimized coefficients of each metasurface are provided in the Supplementary Material. Once the phase for each metasurface was determined, the spacings between the surfaces were varied to obtain the desired intermediate zoom states. As labeled in Figure \ref{fig:Schematic}a, the distance between the compensator and variator optics is denoted $d_1$ and the distance between the variator and the focusing optics is denoted $d_2$. The distance between the focusing lens and the image plane, denoted $d_3$, was held constant at 15 mm while $d_1$ and $d_2$ were allowed to vary. 

\begin{figure*}[!htb]
\centering\includegraphics[width=16cm]{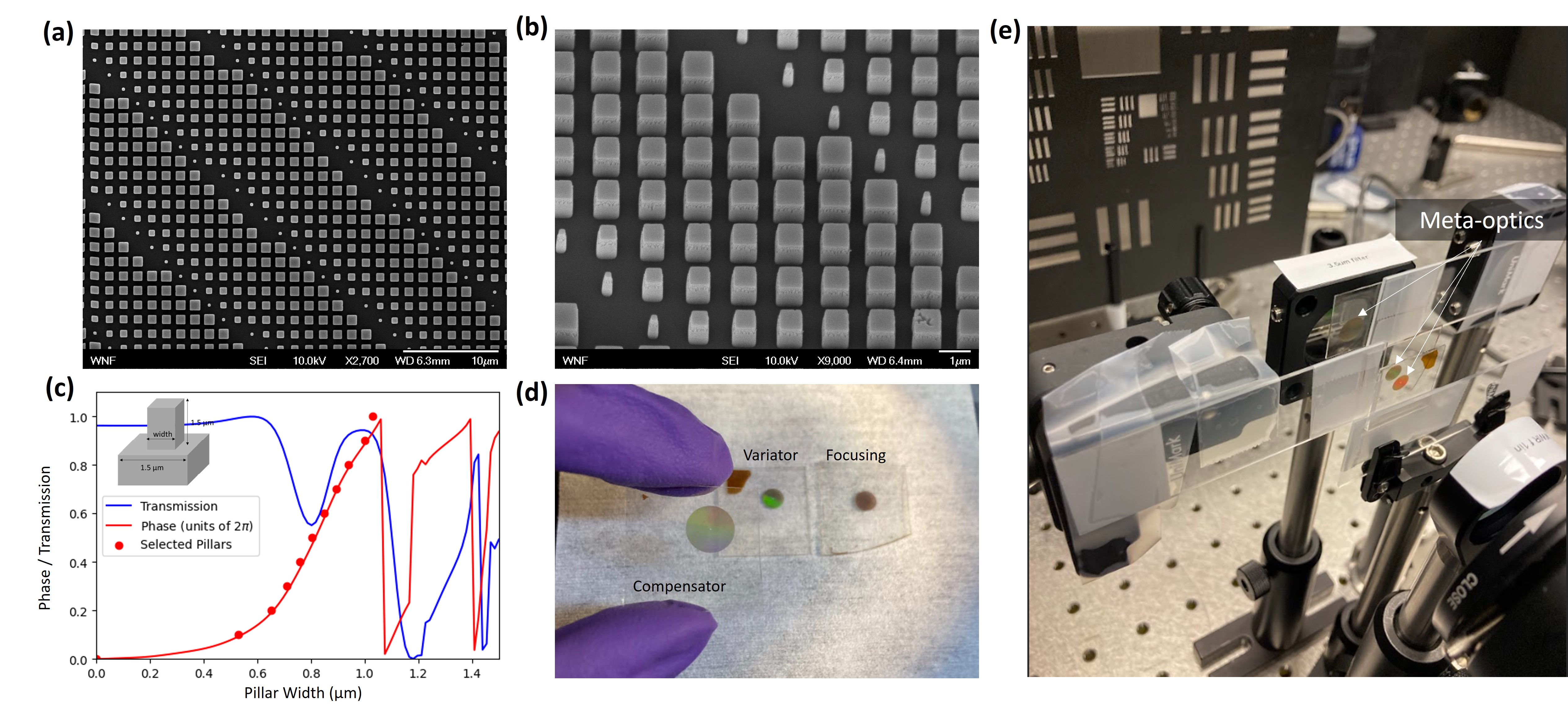}
\caption{\label{fig:MetaOpticsDesign} Metasurface design and fabrication. (a) Top-down SEM image of a fabricated optic. (b) SEM image of a fabricated optic taken at a slightly oblique (10$^\circ$) angle, illustrating fabrication quality. (c) The simulated phase (red) and transmission (blue) of the unit cell. The widths selected for the pillar library are denoted by red circles. (d) A photograph of the fabricated optics. A person is holding the compensator optic in the foreground while the variator (left) and focusing (right) optics are resting on the table in the background. (d) A photograph of the meta-optics in the experiment setup.}
\end{figure*}

\section{Meta-Optics Design and Fabrication}

The meta-optics are physically realized as crystalline silicon pillars on sapphire substrate (commercially available from UniversityWafer). While silicon is transparent at MWIR wavelengths, bulk silicon incurs significant Fresnel reflections due to its high index of refraction (n = 3.5) as compared to the air (n = 1) \cite{Cox58}. Therefore, to provide increased transmission efficiency in the multi-layer system, we chose sapphire (n = 1.7) substrate. Transmission through the sapphire substrate is expected to be 93\%. The silicon pillars are 1.5 $\mu$m tall arranged on a lattice with periodicity 1.5 $\mu$m, and the sapphire substrate thickness is 460 $\mu$m. Using rigorous coupled wave analysis \cite{LiuV12}, we simulated the phase and transmission of silicon pillars as a function of pillar width as shown in Figure \ref{fig:MetaOpticsDesign}c. From these results, we selected ten pillars with widths ranging from 500 nm to 1.0 $\mu$m to comprise the pillar library.  

The meta-optics were fabricated using electron beam lithography and inductively coupled plasma etching. All three meta-opics were fabricated from a single 1.5 $\mu$m thick silicon-on-sapphire wafer that was diced into chips. Each sample was cleaned in acetone and isopropyl alcohol before an oxygen plasma treatment to promote resist adhesion. Next, we spin-coated ZEP-520A resist and exposed the resist by electron beam lithography. After writing the pattern, we developed the resist in amyl acetate followed by a short oxygen plasma treatment to remove residue. In order to achieve a large etching depth of 1.5 $\mu$m, we deposited a hard mask for etching. Specifically, we deposited 70 nm of alumina on top of the patterned samples via electron beam evaporation. To form the hard mask, we lift-off the resist with N-Methyl-2-pyrrolidone. Finally, the silicon was etched via reactive ion etching with SF6 and C4F8 gas chemistry.

Top-down and oblique scanning electron microscope (SEM) images of the fabricated optics are shown in Figure \ref{fig:MetaOpticsDesign}a and b, respectively, illustrating the fabrication quality. A photograph of the fabricated optics is shown in Figure \ref{fig:MetaOpticsDesign}d. Except for the meta-optic area, the thin silicon layer has been etched away, revealing the transparent substrate. The surface of the optic, while appearing opaque brown under visible illumination, is transparent under infrared illumination.

\section{Experiment Imaging Results}

The imaging quality of the meta-optical system was characterized in a simple experimental setup consisting of an illumination source and target object, a narrowband filter, the meta-optics, relay optics, and a camera sensor. A photograph of the fabricated optics in the measurement setup is shown in Figure \ref{fig:MetaOpticsDesign}e. As the source of MWIR illumination, we use a hot plate (Torrey Pines Scientific HP50) set to 130$^\circ$C. This provides broadband infrared radiation that we filter to a narrower band near the design wavelength using a narrowband filter (Thorlabs FB3500-500, with 3.5 $\mu$m center wavelength and FWHM bandwidth of 500 nm). A matte black anodized aluminum target projecting the USAF 1951 resolution chart is placed just after the hot plate to form the imaging object. In order to demonstrate imaging up to $50^\circ$ FoV, we place the target at a distance $D$ from the first meta-optic such that the edges of the target are $25^\circ$ from normal incidence. For our target which is 12.5 cm wide, this corresponds to $D = 13.4$ cm in imaging experiments. Images were collected on a FLIR A6751 MWIR InSb camera sensor with 640 $\times$ 512 pixels at 15 $\mu$m per pixel resolution. Due to the camera sensor being located further than the designated 15 mm back focal length inside the protective camera housing, it was necessary to use relay optics to re-image the meta-optics image onto the sensor. For this purpose, we used an F/1 MWIR-coated, germanium plano-convex singlet lens (Edmund Optics \#69-649, 25 mm focal length) with an F/2.5 compound refractive lens assembly (FLIR \#4218538, 25 mm focal length). To ensure that the relay optics do not restrict measurement of the meta-optics, we ensure that the numerical aperture (NA) of the relay optics  ($NA > 0.45$) exceeds the maximum NA of the meta-optic zoom system ($NA = 0.42$). The relay optics additionally aid in precise alignment of the system; details of the alignment procedure are provided in the Supplementary Material.

\begin{figure*}
\centering\includegraphics[width=16cm]{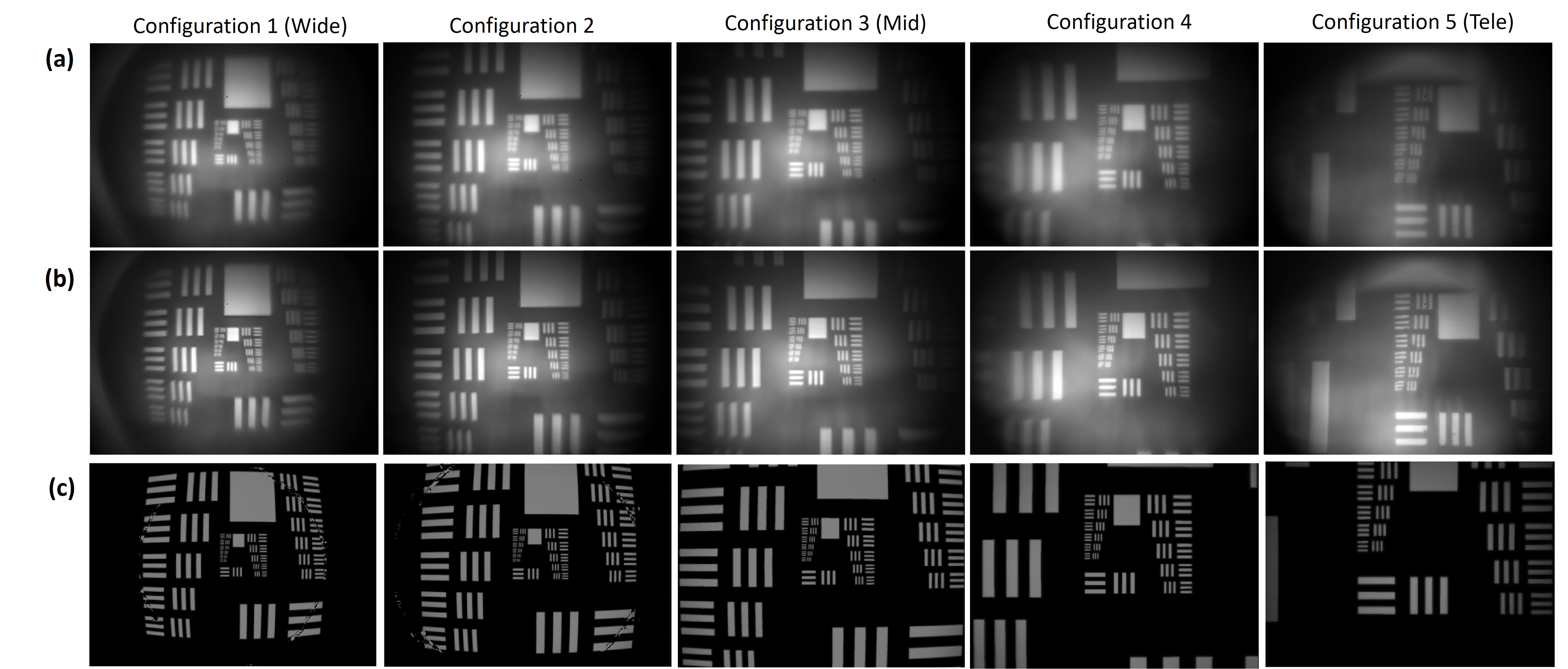}
\caption{\label{fig:Imaging} Simulated and experimental imaging results. (a) Experiment imaging results for Configurations 1 (left) through 5 (right). The imaging object, a hot plate and USAF resolution chart, is placed such that the edges of the target cover 50$^\circ$ FoV with respect to the meta-optics. (b) Experimental results that have been computationally enhanced with sharpening and denoising. (c) Simulated imaging results via ray tracing.  }
\end{figure*}

The simulated and experimental imaging results for all five configurations are shown in Figure \ref{fig:Imaging}, with the USAF resolution target positioned such that the horizontal edges of the imaging target correspond to 50$^\circ$ full FoV. To correct for noise due to ambient radiation on the sensor, we performed a flat field correction on all experimental results by collecting an image with the object blocked and subtracting that background image from the signal; the resulting image captures are shown in Figure \ref{fig:Imaging}a. We observe good imaging quality and close agreement with the simulated results shown in Figure \ref{fig:Imaging}c. However, the somewhat broad linewidth of the illumination source (approximately 500 nm FWHM with the narrowband filter) introduces chromatic aberrations that reduce the sharpness of the image as compared to a single-wavelength measurement. To mitigate this effect, simple computational postprocessing techniques can be introduced to improve the image quality. In Figure \ref{fig:Imaging}b, we show the experiment imaging results after postprocessing with a Gaussian kernel sharpening filter and bm3d denoising algorithm \cite{Maki20}. While there still exists some diffuse noise that we attribute primarily to reflections from the narrowband filter, the USAF groups are effectively sharpened for close qualitative agreement with the simulated single-wavelength results in Figure \ref{fig:Imaging}c. Moving from left to right, the system exhibits clear zoom functionality and qualitatively good imaging performance in Configurations 1 through 5, as well as close agreement between the simulation and experiment. 

\begin{figure*}
\centering\includegraphics[width=14cm]{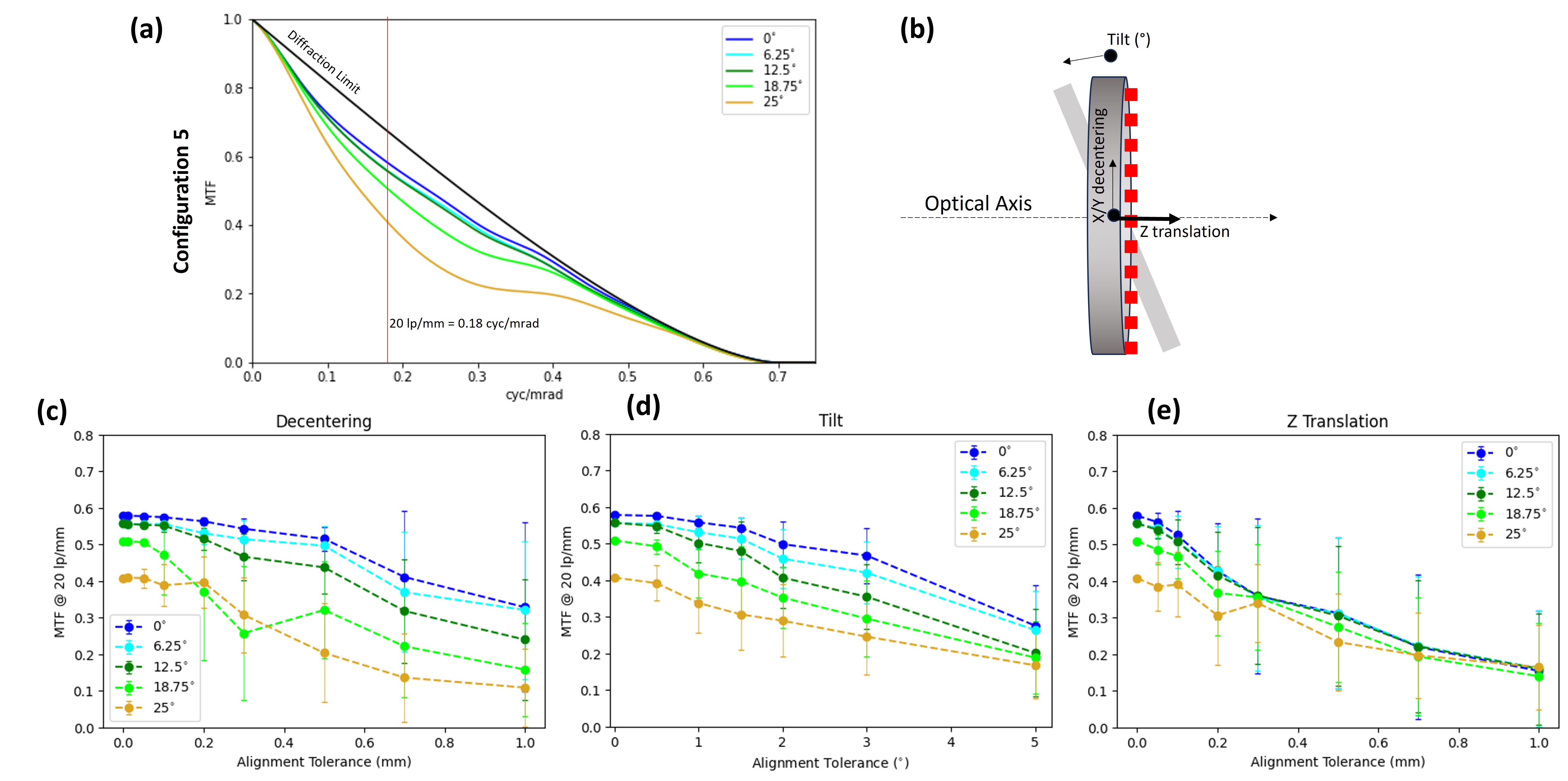}
\caption{\label{fig:Alignment} Tolerance to meta-optic misalignment. (a) Nominal MTF for Configuration 1 for angles of incidence up to 25$^\circ$. The reported MTF is the average over both sagittal and tangential directions. The diffraction limit is plotted in black for comparison. The vertical red line indicates the 20 lp/mm = 0.18 cyc/mrad resolution that was chosen as a metric for the tolerance analysis. (b) Schematic depicting the categories of misalignments that were considered. In the Monte Carlo analysis, all three metasurfaces were misaligned according to the specified tolerances. (c)-(e) Tolerance analysis for the wide configuration. For each field (up to 25$^\circ$), we plot the MTF at 20 lp/mm as a function of allowed misalignment. The misalignment is decentering in (c), tilt in (d), and axial translation in (e). The error bars represent one standard deviation from Monte Carlo analysis consisting of 20 trials.}
\end{figure*}

\section{Discussion}

While this zoom lens system was optimized for single wavelength illumination at 3.4 $\mu$m, we demonstrate high-quality imaging over a wavelength band of approximately 500 nm. In meta-optics (and other forms of diffractive optics) with set phase wraps, the focal length is inversely proportional to the optical wavelength \cite{Huang22}. Therefore, light of different wavelengths does not focus in the same plane, leading to chromatic aberrations. Despite this effect, we demonstrate relatively good imaging quality with the addition of a narrowband filter. While single-wavelength rather than broadband illumination is predicted to improve imaging performance in simulation, further filtering ambient light to a narrower range would also reduce the signal to noise ratio (SNR) due to the reduced light entering the system, unless some active illumination source (such as an LED or laser) is used to illuminate the target. However, such active illumination would be impractical for many applications of this zoom system, such as remote sensing. Therefore, we present imaging results over a broader band, for higher SNR, coupled with computational postprocessing to regain some quality lost due to undesirable chromatic effects. Extension to broader bandwidth operation may be possible using dispersion engineering \cite{Chen20} to match the required phases over a range of wavelengths or applying inverse design techniques \cite{Huang24} to extend the operating wavelength range.

In multi-element optical systems such as this one, misalignment can have deleterious effects on lens performance. To quantify the effects of misalignment on the performance of the system, we conducted a tolerance analysis of Configuration 1 in Zemax at the design wavelength of 3.4 $\mu$m, shown in Figure \ref{fig:Alignment}. The nominal MTF is shown in Figure \ref{fig:Alignment}a. In the absence of misalignment, the imaging performance is nearly diffraction limited at normal incidence and decreases slightly at larger angles of incidence. As a metric of lens system performance, we report the average (sagittal and tangential) MTF at 20 lp/mm resolution as a function of alignment tolerances. As illustrated in Figure \ref{fig:Alignment}b, we specifically consider the effects of translation in Z (axial displacement), translation in X/Y (decentering), and rotation about the X axis (tilt). We varied each category of misalignment separately in this analysis. In detail, we conducted a Monte Carlo analysis comprising 20 trials wherein each element was misaligned by a random amount within the specified tolerance along the horizontal axis. We considered reasonable experimental alignment tolerances of up to 1 mm of translation and up to $5^\circ$ of tilt in each element. In Figure \ref{fig:Alignment}c through \ref{fig:Alignment}e, we report the average MTF value as a function of worsening misalignment for angles of incidence up to 25$^\circ$.  As expected, the average MTF decreases with increasing misalignment, however this decrease is approximately linear. This indicates robustness in the system that we attribute to the relatively small phase gradient of each metasurface. 


Our meta-optics-based zoom lens system achieves a zoom range of 5x, which is sufficient for many applications.  To further increase the zoom range, it would generally require decreasing the NA of the system, which results in reduced imaging quality and field of view \cite{Zhang24}. Similarly to refractive optics, additional layers of optics provide degrees of freedom that can be used to improve performance and achieve additional functionalities, such as increased zoom range. This successful demonstration of a meta-optic triplet system highlights the potential for meta-optics to replace refractive elements in increasingly complex multi-layer and multi-functional imaging systems. In particular, we envision meta-optical systems like this one finding utility in weight-constrained systems requiring zoom functionality and wide field of view, such as airborne remote sensing applications. 

\section*{Funding}
Funding for this work was supported by the federal STTR program.

\section*{Acknowledgements}
Part of this work was conducted at the Washington Nanofabrication Facility / Molecular Analysis Facility, a National Nanotechnology Coordinated Infrastructure (NNCI) site at the University of Washington with partial support from the National Science Foundation via awards NNCI-1542101 and NNCI-2025489.

\section*{Data Availability Statement}

The data that support the findings of this study are available from the corresponding author upon reasonable request.

\section*{Author Contributions}
Z.C. and A.M. conceived and supervised the project. A.W.-S. designed the meta-optic unit cells, led the experimental measurements, and analyzed the data. A.M.J. designed the zoom lens system in Zemax. M.C. fabricated the meta-optics. J.F. and M.C. advised the experimental setup and edited the manuscript. R.J. and T.L.T. contributed to the experimental measurements. A.W.-S. wrote the manuscript with input from all coauthors.

\section*{Disclosures}
A.M. is a co-founder of Tunoptix, which is commercializing similar meta-optics in the visible.
Z.C. and A.M.J. work at CFDRC, which is commercializing infrared imaging systems.



\printbibliography[title={References}]  

\end{document}